\documentclass[prd,showpacs,preprintnumbers,amsmath,amssymb]{revtex4}

\usepackage{graphicx}
\usepackage{dcolumn}
\usepackage{bm}

\def\be{\begin{eqnarray}}
\def\ee{\end{eqnarray}}
\def\bea{\begin{eqnarray}}
\def\eea{\end{eqnarray}}

\def\bT{{\bf b}_\perp}

\def\0T{{\bf 0}_\perp}
\begin{document}


\title{Transverse Force on Transversely Polarized Quarks in Longitudinally Polarized Nucleons}

\author{Manal Abdallah and Matthias Burkardt}
 \affiliation{Department of Physics, New Mexico State University,
Las Cruces, NM 88003-0001, U.S.A.}
\date{\today}

\begin{abstract}
We study the semi-classical interpretation of the $x^3$ and $x^4$ moments of
twist-3 parton distribution functions (PDFs). While no semi-classical 
interpretation for the higher moments of $g_T(x)$ and $e(x)$ was find,
the $x^3$ moment of the chirally odd spin-dependent twist-3 PDF
$h_L^3(x)$ can be related to the longitudinal gradient of the transverse force on transversely polarized quarks in longitudinally polarized nucleons
in a DIS experiment.
We discuss how this result relates to the 
torque acting on a quark in the same experiment. This has further implications for comparisons between tha Jaffe-Manohar and the Ji 
decompositions of the nucleon spin.
\end{abstract}

\maketitle
\section{Introduction}
In the Bjorken limit, cross sections in Deep-Inelastic Scattering (DIS) are usually dominated by 
twist-2 Parton Distribution Functions (PDFs) which have a simple physical interpretation as number densities of quarks with a certain momentum fraction $x$ and
polarization relative to that of the nucleon spin. For example, $f^q(x)$ represents quarks carrying momentum fraction $x$, while $g_1^q(x)$ counts quarks plus anti-quarks with spin in the same direction of the nucleon spin (for a longitudinally polarized nucleon) minus those with spins opposite to the nucleon spin.
The chirally odd twist-2 PDF $h_1^q(x)$ counts quarks (minus antiquarks) with transversity in the same direction
as the nucleon spin for a transversely polarized nucleon minus opposite to that direction.
For higher twist PDFs, whose contribution to cross sections is usually suppressed in the Bjorken limit, no
such simple interpretation in terms of number densities or differences between number densities exists.

In polarized DIS from a transversely polarized target, the contribution from the leading twist PDF $g_1(x)$
is suppressed and even in the Bjorken limit the twist-3 PDF $g_T(x)$ contributes equally
to the longitudinal-transverse double-spin asymmetry. This allows a clean experimental extraction without
contamination from $\frac{1}{Q}$ corrections to the leading twist PDFs.
Likewise, the longitudinal-transverse double-spin asymmetry in polarized Drell-Yan allows accessing the twist-3 PDF $h_L(x)$ \cite{JaffeJi}.

Given that twist-3 PDFs can be measured raises the question what can be learned from these functions. Using the (free) equations of motion, one can identify the so-called Wandzura-Wilczek (WW) part of twist-3 PDFs which is related to the corresponding twist-2 PDF. The remaining part of twist-3 PDFs
is the most interesting as it contains quark-gluon correlations.

In Ref. \cite{mb:force} it was shown that the $x^2$ moment of the twist-3
part of $g_T(x)$ has a semi-classical interpretation as the average transverse force that acts on an unpolarized quark in a transversely polarized nucleon in DIS. Likewise,
the $x^2$ moment of the twist-3 part of the scalar PDF $e(x)$ has an interpretation as
the transverse force acting on a transversely polarized quark in an
unpolarized nucleon. Although these $x^2$ moments only provide information about the local forces, these are the same forces that when integrated along the trajectories of the ejected quark give rise to the Sivers and
Boer-Mulders funtions respectively that describe single-spin asymmetries
in transverse momentum dependent parton distributions.

The $x^2$ moment of the twist-3 part of the chirally odd PDF $h_L(x)$ vanishes
identically. Intuitively, this result can be understood since if its $x^2$
moment were nonzero, it would describe the transverse force on a transversely
polarized quark in a longitudinally polarized nucleon - which would violate parity.

In this work we consider the $x^3$ and $x^4$ moments of these PDFs in an attempt to extend these semiclassical interpretations to higher moments.

\section{Moment Analysis}
The chirally even spin-dependent PDFs are defined as
\bea
\int\frac{d\lambda}{4\pi}e^{i\lambda x} \langle PS|\bar{q}(q)\gamma^\mu
\gamma_5q(\lambda n)|_{Q^2}|PS\rangle = g_1(x,Q^2)p^\mu ({\vec S}\cdot {\vec n})+g_T(x,Q^2)S_\perp^\mu +M^2g_3(x,Q^2)n^\mu ({\vec S}\cdot {\vec n})
\eea
where $p^\mu$ and $n^\mu$ are light-like vectors along the '-' and '+' light-cone direction with $p\cdot n=1$. In the following, for simplicity, the
$Q^2$ dependence will not explicitly written.
$g_T(x)$ can be decomposed as
\be
g_T(x)=g_1(x)+g_2(x)=g_1(x)+g_2^{WW}(x)+\bar{g}_2(x),
\ee
where the Wandzura-Wilczek contribution reads
\be
g_2^{WW}(x)=-g_1(x)+\int_x^1\frac{dy}{y}g_1(y).
\ee
These relations, together with explicit expressions for the
quark-gluon correlations embodied in $\bar{g}_2(x)$ can be derived for each
moment by using the equations of motion
\footnote{This is valid for matrix elements of these operators. However, to
keep notation simple, these matrix elements will not be written at intermediate steps.}
\be
\left[i\gamma^+D_++i\gamma^-D_--i\gamma^xD^x-i\gamma^yD^y\right]q=0,
\ee
to eliminate terms involving the 'bad' component $q_-\equiv \frac{1}{2}
\gamma^+\gamma^-q$ of the field operators. Here we set the quark mass $m=0$for simplicity. For the $x^2$ moment this procedure results in
\be
\frac{3}{2}\bar{q}\gamma^x\gamma_5 D_-^2q
= \frac{1}{2}\bar{q}\left[\gamma^+\gamma_5 \left(D_-D^x +D^xD_-\right)
+ \gamma^x\gamma_5 D_-^2\right]q
+\frac{i}{2}\bar{q}\gamma^+\left[D_-,D^y\right]q,
\label{eq:D-D-d}
\ee
where the first term on the r.h.s. is twist 2 and represented by the $x^2$ moment
of $g_2^{WW}$. The matrix elements of the second term $\frac{1}{2}\bar{q}\gamma^+gG^{+y}q$ have the semi-classical interpretation of the color Lorentz force in the $y$ direction. Repeating the same steps for the
$x^3$ moment results in
\be
\bar{q}\gamma^x\gamma^5D_-^3q = \frac{1}{4}\bar{q}\left[\gamma^x\gamma^5D_-^3
+\gamma^+\gamma^5\left(D_-^2D^x+D_-D^xD_-+X^xD_-^2\right)\right]q
+\frac{1}{8}\bar{q}\gamma^+\gamma^5\left[D_-,\left[D_-,D^x\right]\right]q
+\frac{3i}{8}\bar{q}\gamma^+\bar{q}\gamma^+\left[D_-^2,D^y\right]q,
\ee
where again the first term is completely symmetric in its Lorentz indices and is represented by $\int dx\,g_2^{WW}(x)x^3$. The second term has the
semi-classical interpretation as a force gradient, but we were not able to identify a simple semi-classical interpretation for the third term. The situation is even more complex for the $x^4$ moment.

In the case of the scalar twist 3 distribution $e(x)$ the situation is similar. The twist-3 part of its $x^2$ moment 
\be
\int dx e^{(3)}(x)= \frac{1}{4M}\langle P|\bar{q}\sigma^{+i}gG^{+i}q|P\rangle
\ee
has the semi-classical interpretation of a transverse force on a transversely polarized quark in an unpolarized nucleon. For the $x^3$ moment one also finds a term that can be
interpreted as a force gradient, but there is another term that does not have a simple 
interpretation either. 

In the case of the chirally odd spin-dependent twist 3 distribution $h_L(x)$ the situation is different.  Its $x^2$ moment only yields a twist 2 term \cite{JaffeJi}
\be
\int dx h_L(x)x^2 = \frac{1}{2}\int dx h_1(x)x^2
\ee
For the $x^3$ moment one finds 
\be
2M\int dx\, x^3 h_L^{(3)}(x) =-\frac{i}{{P^+}^3}\frac{1}{6}
\left\langle P,S\left|\bar{q}\gamma^+i\gamma^5\left\{
\gamma^x\left[D_-,gG^{+x}\right] +\gamma^y\left[D_-,gG^{+y}\right]\right\}q
\right|P,S\right\rangle\label{eq:hL}.
\ee
Due to the Dirac matrix $\sigma^{+i}\gamma_5 =i\gamma^+\gamma^i\gamma_5$,
the matrix element projects out the quark transversity asymmetry. $G^{+i}$
represents the transverse color Lorentz force acting on a quark moving with
the velocity of light in the $-\hat{z}$ direction. The right hand side of
Eq. (\ref{eq:hL}) thus describes the average longitudinal gradient of the
transverse force that acts on transversely polarized quarks.
Although that force itself must vanish due to PT invariance, its gradient
$[D_-,G^{+i}]$ is in general nonzero and its sign provides insights about
how the color magnetic field of the nucleon is correlated with its spin.

\begin{figure}
\unitlength1.cm
\begin{picture}(10,8)(-1,12.7)
\includegraphics{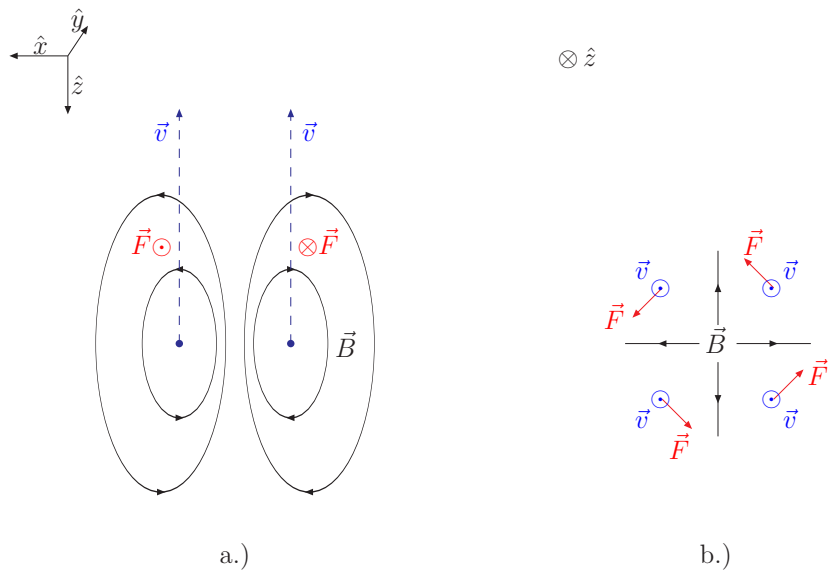}
\end{picture}
\caption{Illustration of the torque acting on a positron moving in the $-\hat{z}$ direction
through a magnetic dipole field caused by an electron polarized in the $+\hat{z}$ direction. 
a.) side view; b.) top view.
In this example the $\hat{z}$ component of the torque is negative as the positron is ejected.
}
\label{fig:dipole}
\end{figure}

To illustrate how this information is embodied in Eq. (\ref{eq:hL}),
consider a valence quark with transversity in the $+\hat{x}$ direction.
As explained in Ref. \cite{HANNAFIOUS}, and confirmed in Lattice QCD
calculations \cite{BM:lattice}, its distribution in the transverse plane
is shifted into the $+\hat{y}$ direction. Suppose the color magnetic
field due to the spectators is oriented as shown in Fig.~\ref{fig:dipole}.
In this example the gradient of the color Lorentz force would thus
point on average in the $-\hat{x}$ direction. If the color magnetic
field has opposite orientation, that force would on average point in the
$+\hat{x}$ direction. Measuring (or calculating in Lattice QCD)
the $x^3$ moment of $h_L^{(3)}$ thus allows one to probe the orientation of
color-magnetic forces in the nucleon.

\section{Discussion}
We find that the $x^3$ moment of the twist-3 chirally odd spin-dependent PDF $h^3_L(x)$ embodies information on the (longitudinal) force gradient acting on a transversely polarized quark in a longitudinally polarized quark as the quark leaves the target in a DIS experiment.
The local force in the same context vanishes due to an ensemble average between quarks starting at the 'bottom' of the nucleon and those that start at the 'top' (bottom here refers to the side at which the virtual photon enters the nucleon.

In Ref. \cite{Burkardt:2012sd}, we explained that the difference between
the definition of quark orbital angular momentum in the nucleon given
by Jaffe-Manohar \cite{g2} and that given by Ji \cite{Ji:PRL,Ji:PRD}
can be expressed in terms of the change in orbital angular momentum as
the quark leaves the target. That change is due to the torque from the
color forces and can be expressed in terms of the matrix element
\begin{eqnarray}
{\cal L}_{JM}^q \! -\! L_{Ji}^q & \!\!\!\! = \!\!\!\!\!\! &
\frac{\left\langle P,S_\parallel\left|\int d^2{\bf r}_\perp
\bar{q}(0^-,{\bf r}_\perp)\gamma^+\int_0^\infty dr^-\left[x F^{+y}(r^-,{\bf r}_\perp)
-y F^{+x}(r^-,{\bf r}_\perp)\right]
q(0^-,{\bf r}_\perp)
\right|PS_\parallel\right\rangle 
}{\left\langle P,S_\parallel |PS_\parallel\right\rangle} \nonumber \\
& &
\label{eq:torque}
\end{eqnarray}
Knowledge about the color-Lorentz forces from (\ref{eq:hL}) also provides
clues about the effects owing to the FSI torque entering
Eq.~(\ref{eq:torque}): The local matrix element
\be
\frac{\left\langle P,S_\parallel\left|\int d^2{\bf r}_\perp
\bar{q}(0^-,{\bf r}_\perp)\gamma^+\left[x F^{+y}(0^-,{\bf r}_\perp)
-y F^{+x}(0^-,{\bf r}_\perp)\right]
q(0^-,{\bf r}_\perp)
\right|PS_\parallel\right\rangle 
}{\left\langle P,S_\parallel |PS_\parallel\right\rangle}\ee
vanishes due to PT invariance, which in more physical terms arises from
a `top' versus `bottom' cancellation when averaging over the nucleon volume
(the volume averaging is implicit due to taking the matrix element in a
plane wave state). In neither Eq. (\ref{eq:torque}) nor the matrix element of the
'torque gradient'
\be
\frac{\left\langle P,S_\parallel\left|\int d^2{\bf r}_\perp
\bar{q}(0^-,{\bf r}_\perp)\left[x D_-F^{+y}(0^-,{\bf r}_\perp)
-y D_-F^{+x}(0^-,{\bf r}_\perp)\right]
q(0^-,{\bf r}_\perp)
\right|PS_\parallel\right\rangle 
}{\left\langle P,S_\parallel |PS_\parallel\right\rangle}
\label{eq:torquegrad}
\ee
vanish as illustrated in Fig. 1.
In an ensemble average, all positrons
ejected in the $-\hat{z}$ direction pass through the region of outward pointing radial
magnetic field component, but only those originating in the bottom portion also move through
regions of inward pointing radial component, i.e. for positrons ejected in the $-\hat{z}$ direction
the regions of outward pointing radial component dominate. For the torque gradient a similar argument results in a non-vanishing ensemble average.
Qualitatively, a similar argument applies for the torque acting on a quark moving through the color-magnetic field caused by the spectators.

The similarity between the matrix elements of the transverse force and torque also points towards one potential application of this work. In Ref. \cite{HANNAFIOUS} it observed that there seem to be a universal sideways deformation for the impact parameter distribution of transversely polarized quark. That result was based on studying a variety of models and confirmed by lattice QCD calculations. That deformation should be independent on the longitudinal polarization due to PT. Based on these works one knows the direction into which the distribution of quarks with a given transverse polarization is shifted as well as the approximate magnitude of the shift. The sign of the matrix element
for the force gradient, in combination with the above shift, should thus 
provide information about the sign and approximate magnitude of the
torque gradient (\ref{eq:torquegrad}). If one makes the further assumption that the integrand in the integral over $dr^-$ in Eq. (\ref{eq:torque})
does not fluctuate in sign as an function of $r^-$ it would allow to 
predict the sign of ${\cal L}^q_{JM}-L^q_{Ji}$. This would be very valuable when comparing the two corresponding decompositions of the nucleon spin.

{\bf Acknowledgements:}

This work was partially supported by the DOE under grant number 
DE-FG03-95ER40965. 

\end{document}